\documentclass{jpp}
\usepackage{graphicx}
\usepackage{epstopdf, epsfig}
\usepackage{hyperref}

\usepackage{amsfonts}
\usepackage{times}
\usepackage{graphicx}
\usepackage{natbib}

\renewcommand{\vec}[1]{\mbox{\boldmath $#1$}}

\def \Om  {{\it \Omega}}
\def \rin {r_{\rm in}}
\def \mur {r_{\rm in}}
\def \Rey {\ensuremath{\rm{Re}}}
\def \Ha {\ensuremath{\rm{Ha}}}
\def \Hamin {\ensuremath{\rm{Ha_{min}}}}
\def \Pm {\ensuremath{\rm{Pm}}}
\def \Rm {\ensuremath{\rm{Rm}}}
\def \Mm {\ensuremath{\rm{Mm}}}

\def\beg{\begin{equation}}
\def\ende{\end{equation}}
\newcommand{\gsim}{\lower.7ex\hbox{$\;\stackrel{\textstyle>}{\sim}\;$}}
\newcommand{\lsim}{\lower.7ex\hbox{$\;\stackrel{\textstyle<}{\sim}\;$}}
\renewcommand{\vec}[1]{\mbox{\boldmath $#1$}}
\def\curl{{\rm curl}} 
\def\Om{{\it \Omega}}

\def\ara\&a{ Ann. Rev. Astronomy Astrophysics}

\title{Azimuthal magnetorotational instability  with  super-rotation}

\author{G. R\"udiger\aff{1}
  \corresp{\email{gruediger@aip.de}},
  M. Schultz\aff{1}, M. Gellert\aff{1}
 \and F. Stefani\aff{2}}

\affiliation{\aff{1}Leibniz-Institut f\"ur Astrophysik Potsdam, An der Sternwarte 16, D-14482 Potsdam, Germany
\aff{2}Helmholtz-Zentrum Dresden-Rossendorf, PF 510119, 01314 Dresden, Germany}

\begin{document}

\maketitle

\begin{abstract}
It is demonstrated that the azimuthal magnetorotational instability (AMRI)
also works with  radially increasing rotation rates  contrary to the standard magnetorotational instability for axial fields
which { requires} negative shear. The stability against nonaxisymmetric
perturbations of a conducting Taylor-Couette flow with positive shear under
the influence of a toroidal magnetic field  is considered  if the background field between the cylinders is current-free. 
For small magnetic Prandtl number  $\Pm\to 0$ the curves of neutral  stability converge  in 
the Hartmann number/Reynolds number plane approximating the stability curve obtained in the inductionless limit $\Pm=0$. The numerical solutions for $\Pm=0$ indicate the existence of  a lower limit of the  shear rate.  For large $\Pm$ the curves scale with the magnetic Reynolds number  of the outer cylinder but the flow is always stable for magnetic Prandtl number unity as is typical for double-diffusive instabilities.

We are  particularly interested to know  the minimum Hartmann number  for neutral stability. For models with resting or  almost  resting inner cylinder and with perfect-conducting cylinder material   the minimum Hartmann number occurs for a radius ratio of $\rin=0.9$. The  corresponding critical  Reynolds numbers are smaller than  $10^4$.  

\end{abstract}
\keywords  {Astrophysical plasma -- Plasma instabilities}
\section{Introduction}
In recent years instabilities in rotating conducting fluids under the influence of magnetic fields
became of high interest. Especially in view of astrophysical applications  the consideration of
differential rotation $\Om=\Om(R)$ is relevant. It has been  known for a long time that differential rotation
with negative shear (${\rm d}\Om/{\rm d}R<0$, `sub-rotation') becomes unstable under the influence of an axial field \citep{V59,BH91}.
When considered separately, both ingredients of this MagnetoRotational Instability (MRI), i.e. the
axial field and also the  Kepler rotation, are stable. The full system proves to be unstable if and
only if the fluid is sub-rotating.  For ideal fluids the criterion of stability against axisymmetric
perturbations of differential rotation under the presence of {\em toroidal} fields $B_\phi$ reads 
\begin{equation}
 \frac{1}{R^3}\frac{{\rm{d}}}{{\rm{d}}R}(R^2\Om)^2-\frac{R}{\mu_0\rho}
 \frac{{\rm{d}}}{{\rm{d}}R}\left( \frac{B_{\phi}}{R} \right)^2 > 0,
\label{mich} 
\end{equation}
where $\mu_0$ is the magnetic permeability, $\rho$ the mass density and  $R$ the distance from the rotation axis \citep{M54}.

The stability problem of toroidal fields plus differential rotation in 
cylindric geometry has  been studied  by several authors often considering the fluid as ideal and/or the  perturbations as 
axisymmetric  \citep{C61,G62,HG62,C79,F84,BH92,K92,DK93,OP96,TP96,PP05,S06}. In view of possible experiments in the present paper the perturbations  are considered as nonaxisymmetric existing in a diffusive  fluid under the influence of a current-free toroidal magnetic field.

After (\ref{mich}) solid-body rotation and  rotation laws with positive shear  (${\rm d}\Om/{\rm d}R>0$, `super-rotation')
are {\em stabilized} by the magnetic field unless the field strongly increases outwards. The
profiles $B_\phi\propto 1/R$ (current-free between the cylinders) and $B_\phi\propto R$ ($z$-pinch,
homogeneous axial electric-current)  
cannot destabilize rotation  increasing radially
if only axisymmetric perturbations are considered. It has even been demonstrated that for fields with
$B_\phi\propto 1/R$ {\em all} Taylor-Couette flows are stable against axisymmetric perturbations \citep{HS06}. 
The Azimuthal MagnetoRotational Instability (AMRI) of differential rotation plus  azimuthal fields  current-free  between the cylinders is thus basically nonaxisymmetric. 
Standard MRI and AMRI  have in common that in both cases the magnetic background fields are force-free and that the combination of two stable components leads to instability. If the azimuthal  magnetic field is not current-free and axial electric background currents exist we speak about Tayler instability (TI), which even exists without any rotation.

Taylor-Couette flows with positive shear are the prototype of hydrodynamic stability
for moderate Reynolds numbers \citep{W33,SG59}.  For finite, but very large,   Reynolds numbers  the existence of  a linear instability for super-rotating Taylor-Couette flow has been reported recently \citep{D17}. All the  more surprising is the finding that   weak and stable magnetic {\em azimuthal}  fields  can
  destabilize also flows   with radially increasing $\Om$ of rather low Reynolds number  if  nonaxisymmetric perturbations are taken into account
\citep{SK15,RS16}. Even in this particular  combination of two highly stable ingredients
the resulting MHD flow becomes unstable.  We shall also  demonstrate that the radial profile of the fields -- if not too
steep -- does not play an important role for the question of stability or instability.  On the other hand, the new  instability is a double-diffusive phenomenon as it does not appear if the molecular viscosity and the molecular magnetic resistivity are equal.
Contrary to AMRI,   the  standard MRI with  {\em axial} magnetic fields and differential rotation with negative shear  has no counterpart for  positive shear.

The equations of the problem are 
\begin{eqnarray}
 \frac{\partial \vec{U}}{\partial t} + (\vec{U}\cdot \nabla)\vec{U}& =& -\frac{1}{\rho} \nabla P + \nu \Delta \vec{U} 
   + \frac{1}{\mu_0\rho}{\textrm{curl}}\vec{B} \times \vec{B},\nonumber\\
 \frac{\partial \vec{B}}{\partial t}&=& {\textrm{curl}} (\vec{U} \times \vec{B}) + \eta \Delta\vec{B}  
   \label{mhd2}
\end{eqnarray}
with $  {\textrm{div}}\ \vec{U} = {\textrm{div}}\ \vec{B} = 0$ for an incompressible fluid.
$\vec{U}$ is the velocity, $\vec{B}$ the magnetic field vector, $P$ the pressure, $\nu$ the kinematic
viscosity and $\eta$ is the magnetic resistivity. The  basic state in the cylindric system with the
coordinates ($R,\phi,z$) is \mbox{$ U_R=U_z=B_R=B_z=0$} for the poloidal components and 
\mbox{$\Om  = a_\Om  + {b_\Om}/{R^2}$} for the rotation law with the constants 
\begin{eqnarray}
 a_\Om=\frac{\mu-\mur^2}{1-\mur^2}\Om_{\rm in}, \quad\quad\quad\quad
 b_\Om= \frac{1-\mu}{1-\mur^2}\Om_{\rm in} R_{\rm in}^2.
 \label{ab} 
\end{eqnarray}
Here $\mur={R_{\rm in}}/{R_{\rm out}}$ is the ratio of the inner cylinder radius $R_{\rm in}$ and the outer cylinder 
radius $R_{\rm out}$.  $\Om_{\rm in}$ and $\Om_{\rm out}$ are the angular velocities of the inner and outer cylinders, 
respectively. With the definition 
\beg
\mu=\frac{ \Om_{\rm out}}{\Om_{\rm in}}
\label{mu}
\ende
super-rotation is represented by $\mu>1$. 

For the magnetic field the stationary solution is 
\begin{eqnarray}
 B_\phi=a_B R+\frac{b_B}{R}.
 \label{basicB} 
\end{eqnarray}
The radial magnetic profile $B_\phi\propto R$ is due to an applied homogeneous axial electric
current  while $B_\phi\propto 1/R$  is current-free in the fluid.    We define
$\mu_B=B_{\rm out}/B_{\rm in}$. The current-free 
field is then given by $\mu_B=\rin$ and the field in the $z$-pinch is $\mu_B=1/\rin$. Almost always in the present paper a narrow  gap with $\rin=0.9$ is considered.

The dimensionless physical parameters of the system are the magnetic Prandtl number $\Pm$, the 
Hartmann number $\Ha$ and the Reynolds number $\Rey$, i.e. 
\begin{eqnarray}
 {\Pm} = \frac{\nu}{\eta}, \quad\quad\quad
 {\Ha} =\frac{B_{\rm in} D}{\sqrt{\mu_0\rho\nu\eta}},  \quad\quad\quad
 {\Rey} =\frac{\Om_{\rm out} D^2}{\nu}.
\label{pm}
\end{eqnarray}
The magnetic Reynolds number is $\Rm=\Pm\cdot \Rey$. The parameters combine to the magnetic Mach number  
\begin{eqnarray}
{\Mm}=\frac{R_{\rm out}}{D}\frac{\sqrt{\Pm}\Rey}{\Ha},
 \label{mach}
 \end{eqnarray}
 indicating
whether the rotation energy dominates the magnetic energy or not. The parameter $D=R_{\rm out}-R_{\rm in}$ 
is the gap width. The Hartmann number is defined by the magnetic field  on the inner wall where it is maximal. If the Reynolds number is defined with the rotation rate of the outer cylinder, then $\Rey/\mu$ describes the Reynolds number  of the inner cylinder.  $\mu=\infty$ gives a model where the  inner cylinder rests.

The variables  $\vec{U}$, $\vec{B}$ and $P$ are split into mean and fluctuating components, i.e. $\vec{U}=\bar{ \vec{U}}+\vec{u}$, $\vec{B}=\bar{ \vec{B}}+\vec{b}$ and $P=\bar P+p$. The bars from the variables are immediately dropped, so that the upper-case letters $\vec{U}$, $\vec{B}$ and $P$ represent the  background quantities. By developing the disturbances $\vec{u}$, $p$ and $\vec{b}$ into normal modes the solutions of the linearized MHD equations 
\beg
\Big[\vec{u},\vec{b},p\Big]=\Big[\vec{u}(R),\vec{b}(R),p(R)\Big] {\rm e}^{{\rm i}(\omega t+kz+ m\phi)},
\label{fluc}
\ende
are considered for axially unbounded cylinders. Here $k$ is the axial wave number of the perturbation, $m$ its azimuthal wave number and $\omega$ the complex frequency including growth rate as its imaginary part and  a drift  frequency $\omega_{\rm dr}$ as its real part. 
A linear code is used to solve the resulting  set of linearized ordinary differential equations 
for the radial functions of flow, field and pressure  fluctuations. The solutions are
optimized with respect to the Reynolds number for given Hartmann number by varying the wave
number. Only solutions for $m=1$ are here discussed. The hydrodynamic boundary
conditions at the cylinder walls are the rigid ones,  i.e. $u_R=u_\phi=u_z=0$. The cylinders are assumed to be 
perfectly conducting or -- in a few cases -- insulating. For the conducting walls the
fluctuations $\vec{b}$ must fulfill ${\rm d} b_\phi/{\rm d}R + b_\phi/R=b_R=0$ at  $R_{\rm in}$
and $R_{\rm out}$ so that ten boundary conditions exist for the set of ten differential equations. For $B_z=0$ one can easily show that the system is degenerate  under the
transformation $m\to -m$ so that all eigenvalues (for specific $\Rey$ and $\Ha$) are  valid for each 
pair $m=\pm 1$.

\section{Azimuthal magnetorotational instability}\label{sec_amri}
\begin{figure}
  \centerline{\includegraphics[width=0.65\textwidth]{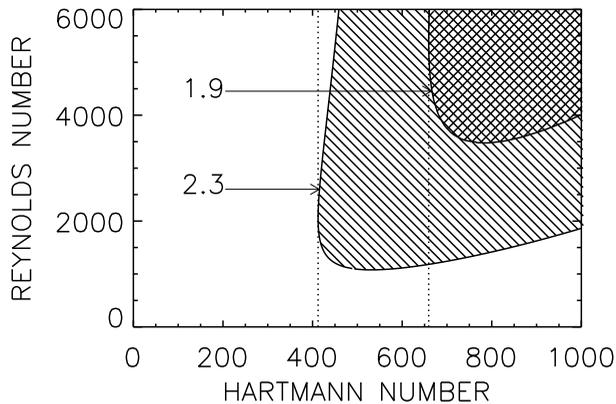}}
  \caption{Instability cones for  two rotation profiles with $\mu=1.9$   
          and $\mu=2.3$. By the vertical dotted lines the minimum Hartmann numbers $\Hamin$ are defined. $\mu_B=\mur=0.9$, $m=\pm 1$,  $\Pm=10^{-5}$. The stability lines are also valid for $\Pm=0$. Perfect-conducting cylinders.}
\label{fig1}
\end{figure}

Consider  rotation laws  where the outer cylinder rotates with higher frequency  than the inner
cylinder.  For the small magnetic Prandtl number of $\Pm=10^{-5}$ Fig. \ref{fig1}  presents the
lines of marginal instability (i.e. for vanishing growth rate) for the two rotation profiles with
$\mu=1.9$ and $\mu=2.3$ for a narrow gap with $\rin=0.9$. The magnetic field between the perfect-conducting cylinders is fixed to the
vacuum-type, $B_\phi\propto 1/R$. The form of the resulting neutral  lines corresponds to the well-known form for  sub-rotation  as a
tilted cone with  both branches having a positive slope \citep{HT10,RK13}.  Absolute   minima of the
Hartmann number and the Reynolds number exist, below which  the rotation law is stable. Above the
minima  the  instability domain is always limited by two critical values of the Hartmann number or
the Reynolds number. For a given supercritical Reynolds number there is a minimum magnetic field
for the instability and there is a maximum magnetic field  destructing the instability. Also the Reynolds number can be too small or too large for a given Hartmann number.  Above the
upper  branch of the instability curve the rotational shear is too strong to support nonaxisymmetric perturbations. Here we are in particular interested in the values of the absolute minimum $\Hamin$ of the Hartmann number for neutral stability in order to discuss  the possibility 
of laboratory experiments. Without magnetic field the flow is stable while the vertical dotted lines in Fig. \ref{fig1} characterize the minimum magnetic field via $\Ha_{\rm min}$. 
\begin{figure}
 \centering
  \includegraphics[width=0.48\textwidth]{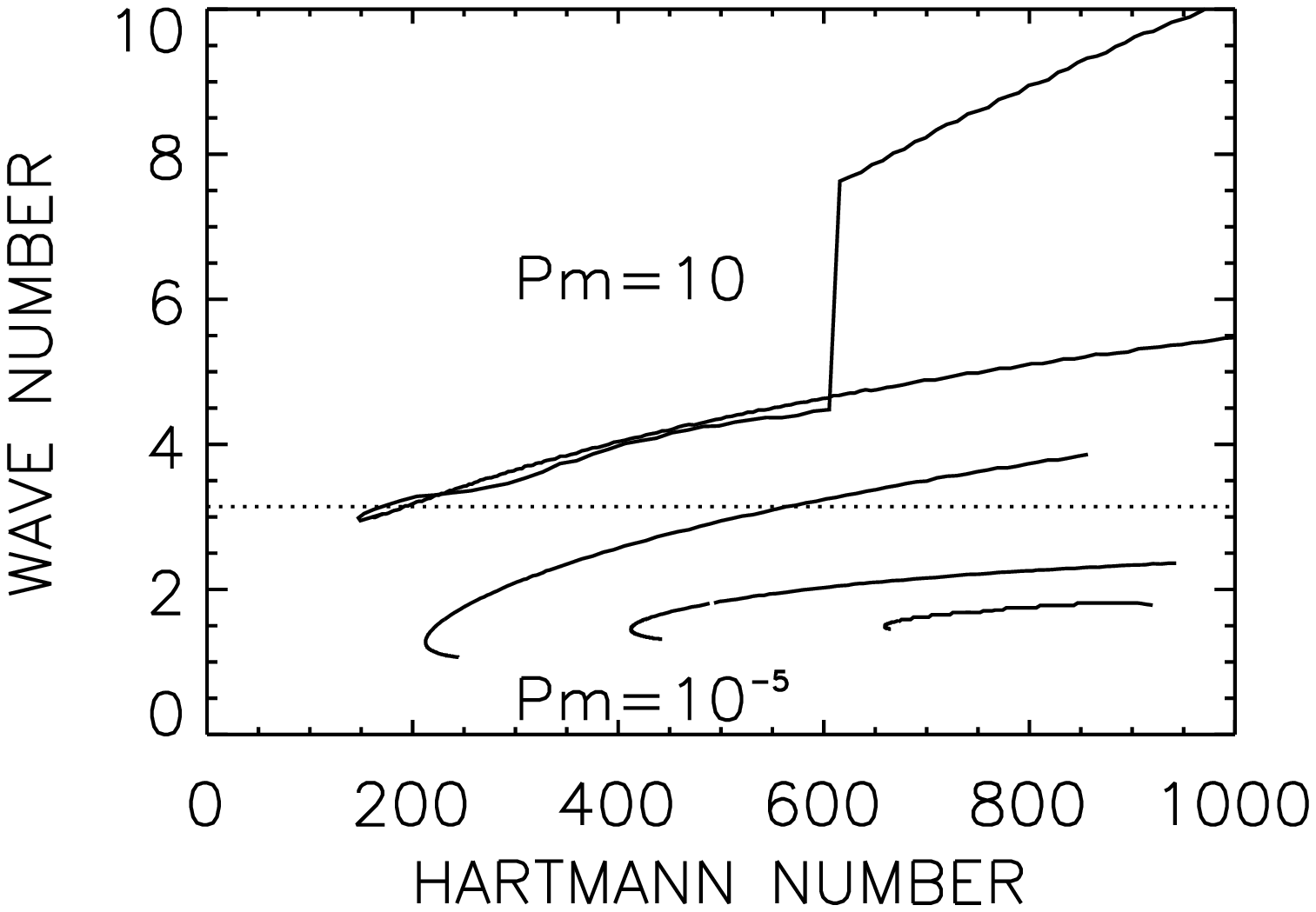}
 \includegraphics[width=0.48\textwidth]{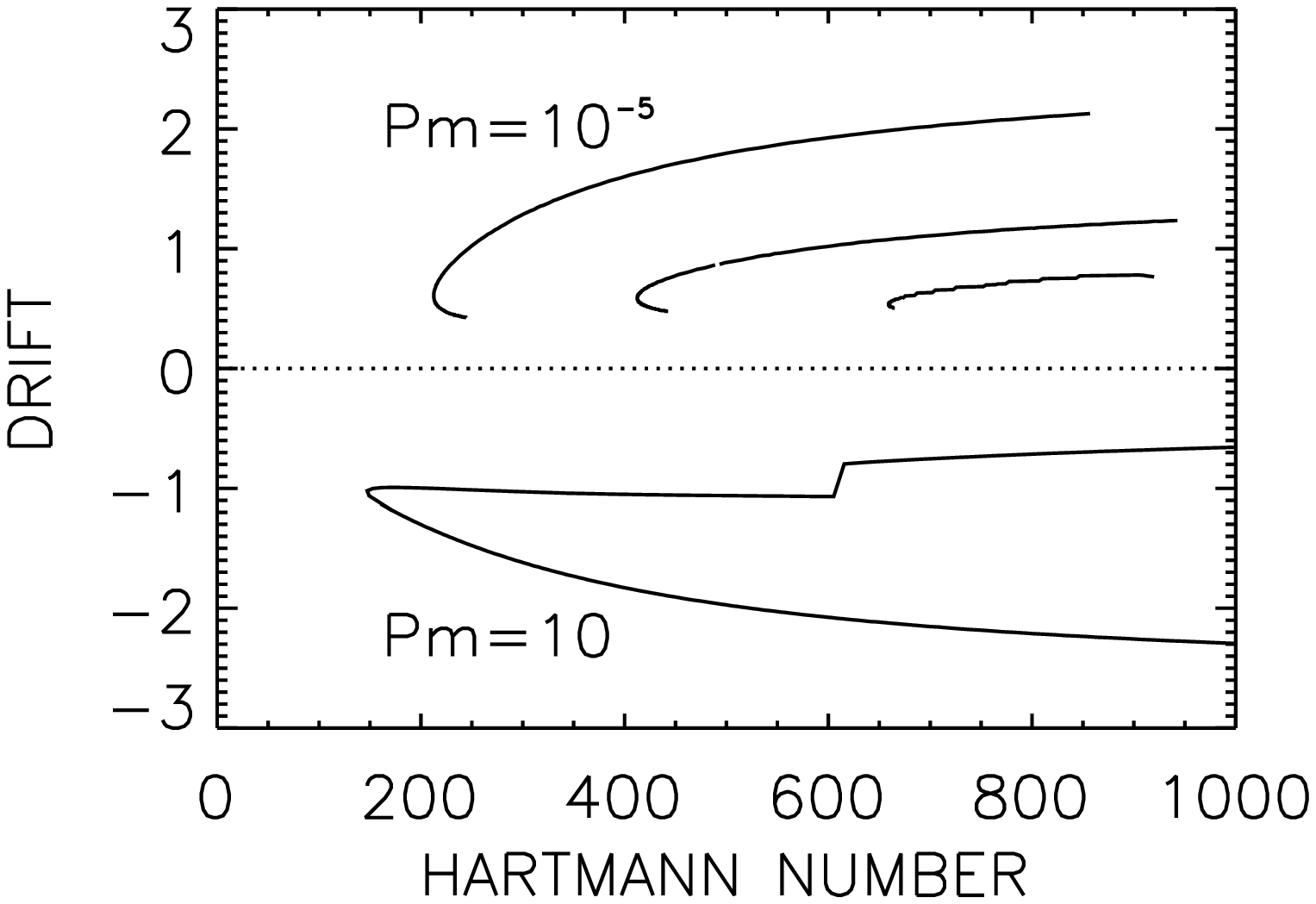}
 \caption{The axial wave number $kD$ (left panel) and  the  drift frequency $\omega_{\rm dr}/\Om_{\rm out}$ (right panel) 
          along the lines of marginal instability for  $\Pm=10^{-5}$ ($\mu=5,  2.3, 1.9$, from left to right) 
          and $\Pm=10$ ($\mu=5$). $\mu_B=\mur=0.9$, $m= \pm 1$. Perfect-conducting cylinders.} 
 \label{fig2} 
\end{figure}

The left panel of Fig. \ref{fig2} gives the optimized axial wave numbers of the instability normalized
with the gap width $D$ along the two branches of instability for small and large $\Pm$. The axial
cell size is $\delta z \simeq \pi D/k$. Cells with $k<\pi$ are thus  prolate while cells with
$k>\pi$ are oblate with respect to the rotation axis. The limit $k=\pi$ for nearly circular  cells is
marked by a horizontal dotted line. The plot demonstrates  that the cell geometry strongly depends
on the magnetic Prandtl number.  For small $\Pm$ also the wave numbers are small hence the cells are
prolate or circular in the ($R,z$) plane. Along the strong-field (lower) branch of the instability cone the axial wave numbers
exceed  those  of the low-field (upper) branch where  the cells are rather  long  in axial
direction. For $\Pm\gg 1$, however, the wave numbers at both branches of the  cone are  larger than
$\pi$ so that    the cells are always very flat.

Also the drift rates possess a strong $\Pm$-dependence. The drift values as given in the right
panel of Fig. \ref{fig2} are the real  parts $\omega_{\rm dr}$ of the frequency $\omega$ of the
Fourier mode  of the instability  normalized with the rotation rate of the {\em outer}  cylinder.
Because of  
\begin{equation}
\frac{ \dot\phi}{\Om_{\rm out}}=-\frac{\omega_{\rm dr}}{m \Om_{\rm out}} 
 \label{phi}
 \end{equation}
  the azimuthal migration $\dot\phi$ has the opposite sign of
$\omega_{\rm dr}$.   For AMRI with negative shear  we always found that the pattern migrates for all
$\Pm$ in positive  $\phi$-direction  \citep{RG14}. For Tayler instability under the influence of
radially increasing rotation  the  situation is more complicated as  the pattern counter-rotates for small $\Pm$
while it co-rotates for $\Pm> 1$.  Figure \ref{fig2} shows similar results.
For small $\Pm$   positive $\omega_{\rm dr}$ occur and the  perturbation pattern indeed  rotates
{\em retrograde}. Large $\Pm$ provide negative drift values hence the pattern migrates in the direction of the 
rotation.  For $\omega_{\rm dr}/ \Om_{\rm out}=-1 $  the pattern  rotates just as the outer cylinder.

\begin{figure}
 \centering
 \includegraphics[width=0.48\textwidth]{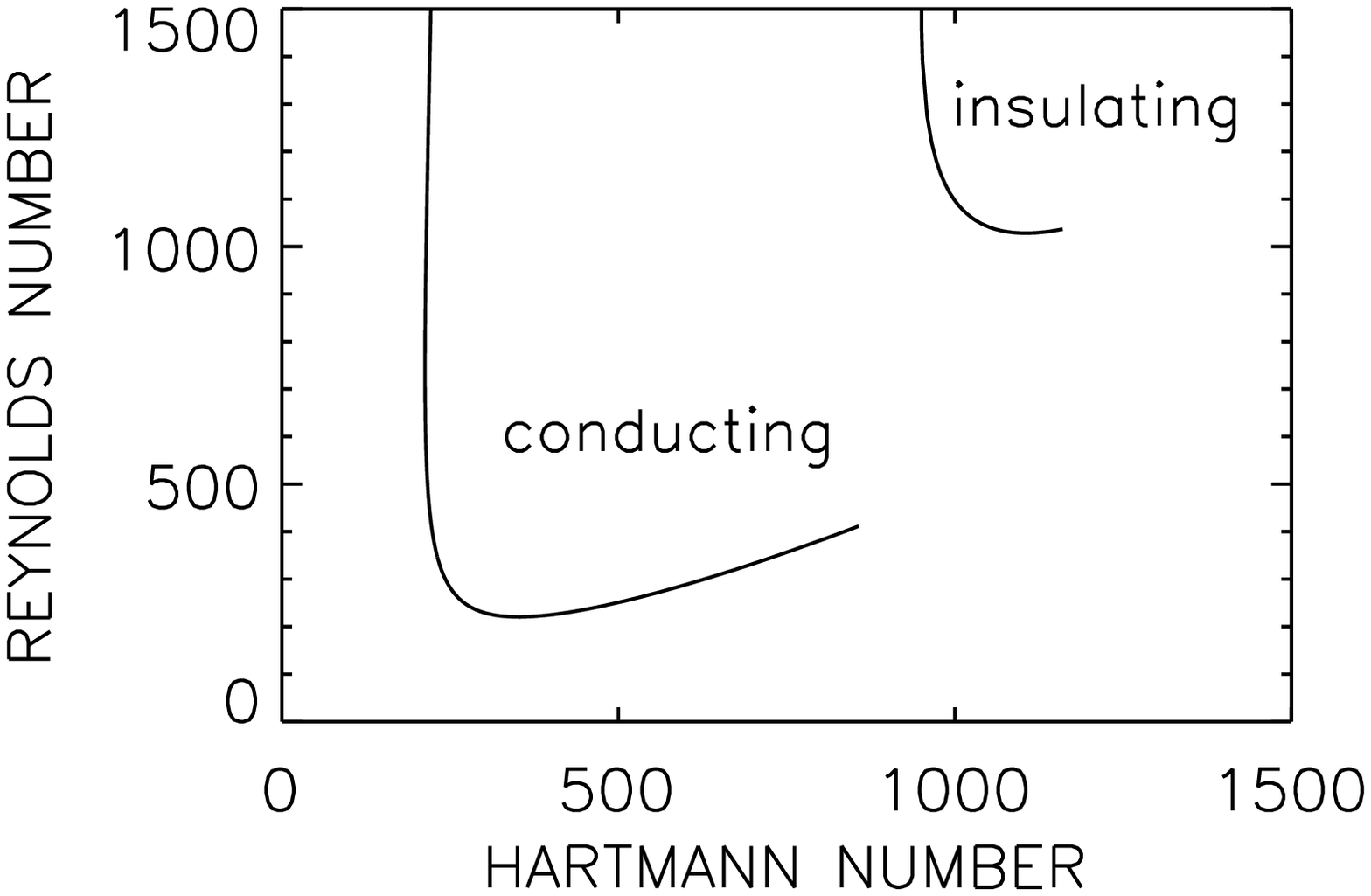}
 \includegraphics[width=0.48\textwidth]{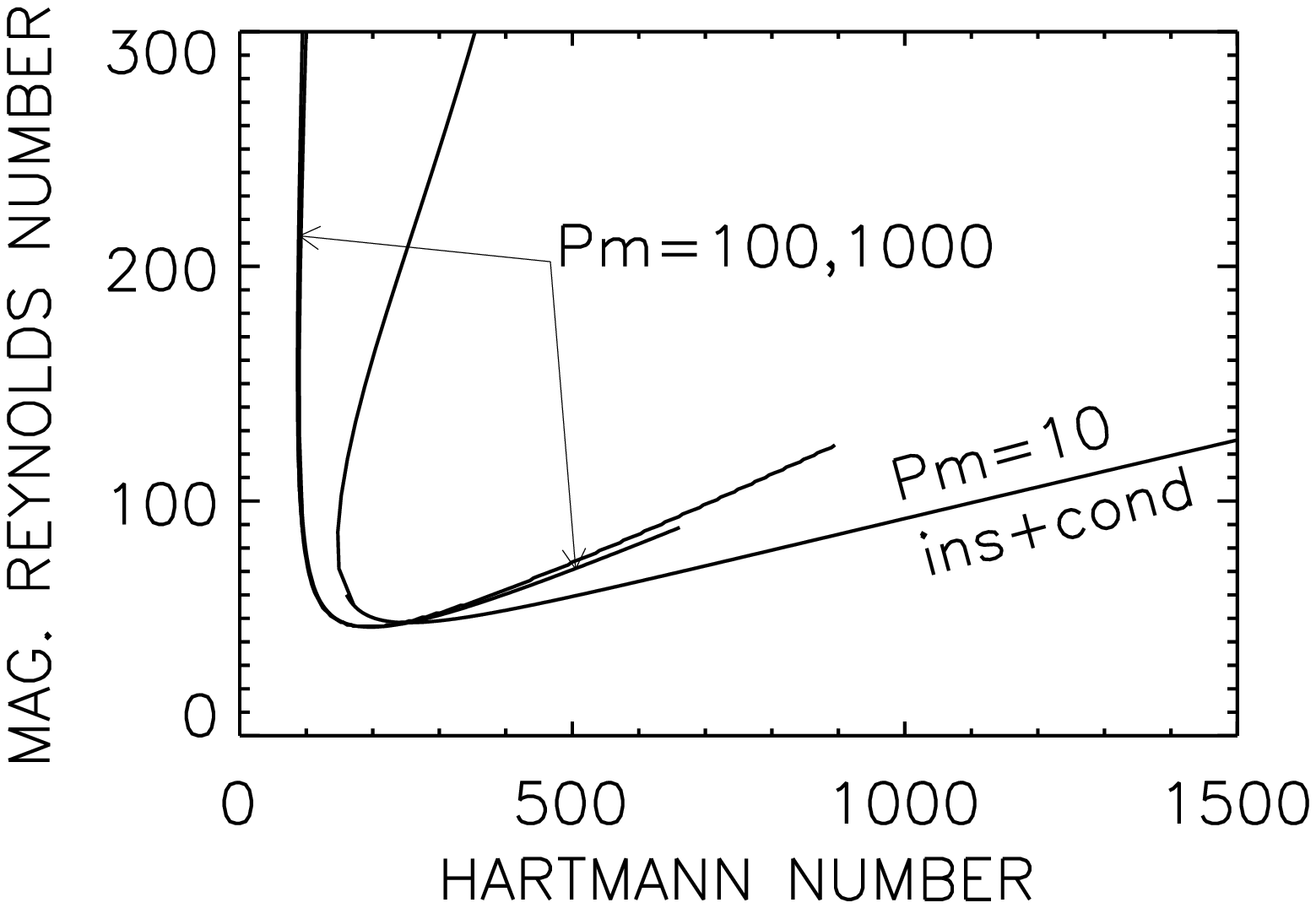} 
 \caption{Stability maps for  $\mu=5$   for perfect-conducting    cylinders for  $\Pm=10^{-5}$ (left) and $\Pm\gg1$ (right).  The lines for $\Pm=100$ and $\Pm=1000$ are almost indistinguishable. The left panel also contains the line of neutral stability for insulating cylinders; the curve for  $\Pm=10$ (right panel) holds for insulating and perfect-conducting boundary conditions.   Solutions for $\Pm= 1$ do not exist.   $\Mm<1$ for all curves. 
       $\mu_B=\mur=0.9$.} 
\label{fig4} 
\end{figure}

To see the influence of the boundary condition the left panel of  Fig. \ref{fig4}   gives
the stability map for $\mu=5$ for the two cases with perfect-conducting walls and insulating
boundary conditions. The small magnetic Prandtl number of $\Pm=10^{-5}$ characterizes liquid sodium. For vacuum boundaries the rotation laws are much more
stable than for perfect-conducting conditions. Note that $\Hamin$  for insulating cylinders exceed the values of $\Hamin$ for perfect-conducting cylinders by a factor of 4. 
Also for AMRI with negative shear the critical Hartmann
numbers and the Reynolds numbers  for vacuum  conditions  strictly  lie above the values for  perfect-conducting  conditions.
One  finds from the Figs. \ref{fig1}--\ref{fig4} the expected  trend with $\mu$ for the
minimum Hartmann  number. For perfect-conducting boundaries and $\Pm=10^{-5}$ it decreases from $\Ha_{\rm min}\simeq 700$ 
for the flat rotation law with $\mu=1.9$ to $\Ha_{\rm min}\simeq 200$  for the steeper rotation law with $\mu=5$.  Below we shall discuss whether the reduction of  $\Ha_{\rm min}$ by growing rotation ratio $\mu$ is continued or not for $\mu\to\infty$, i.e.   for stationary   inner cylinder.

The right panel of Fig. \ref{fig4} demonstrates instability also  for $\Pm\gg 1$.  The stability curves for $\Pm=10$, $\Pm=100$ and $\Pm=1000$ are given  for perfect-conducting boundary conditions in the $\Ha/\Rm$ plane with  $\Rm$ as the  magnetic Reynolds number. Note that in this representation the minimum values for Hartmann number and magnetic Reynolds number do hardly depend on $\Pm$. For  $\Pm\to \infty$ the critical rotation rate  appears to scale with $\Rm$. The curve for $\Pm=10$ simultaneously holds for the boundary conditions of both perfectly conducting and insulating cylinders. For  $\Pm> 1$ the large gaps between these curves  known for   $\Pm<1$ disappear.  The role of the boundary conditions for the excitation of the instability strongly depends on the choice of the magnetic Prandtl number.


\section{Role  of the magnetic Prandtl number}
\begin{figure}
 \centering
 \includegraphics[width=0.48\textwidth]{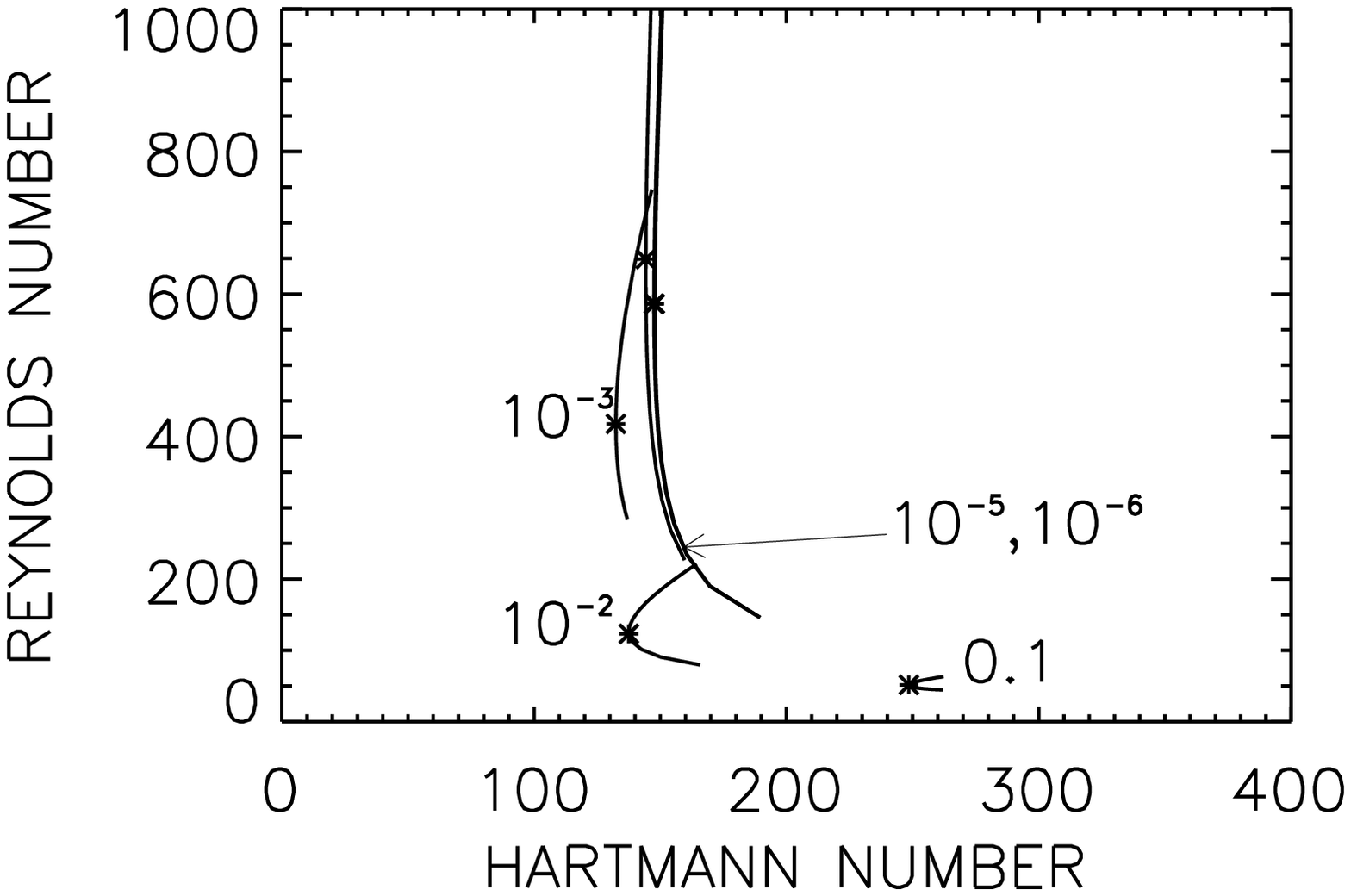}
 \includegraphics[width=0.48\textwidth]{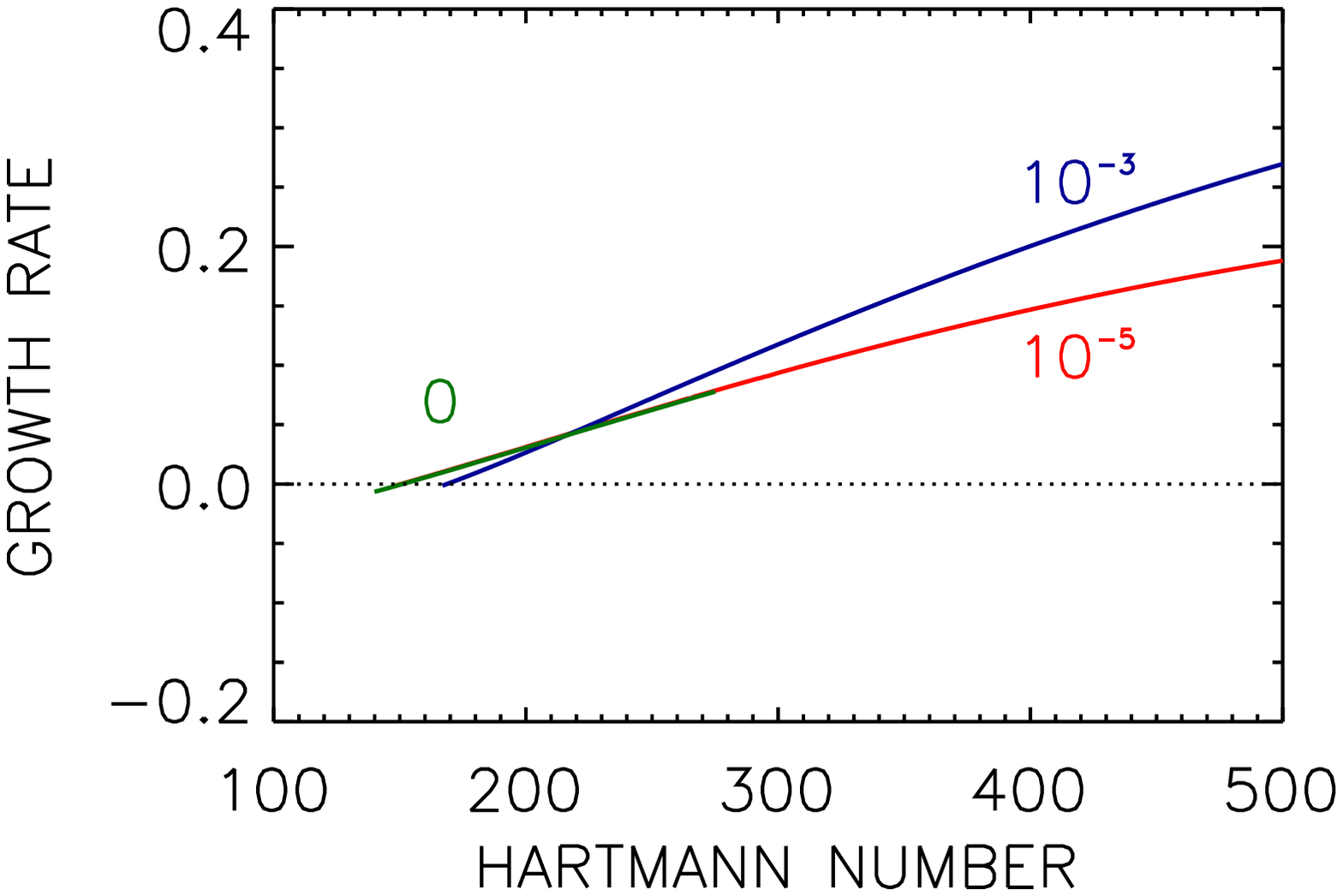}
 \caption{Left: Lines of neutral stability in the ($\Ha/\Rey$) plane for various magnetic Prandtl numbers (marked) for almost stationary inner cylinder. For small $\Pm$ the lines coincide. Their minimum Hartmann number is  $\Hamin\simeq 147$. Stars denote the $\Hamin$. Right: Growth rate normalized with $\Om_{\rm out}$ for $\Rey=1000$ and three magnetic Prandtl numbers as indicated. $\mu_B=\mur=0.9$, $\mu=128$. Perfect-conducting cylinders.}
 \label{fig33}
\end{figure}

We should  also be interested to model containers with resting inner cylinder described by $\mu\to\infty$ which our code approximates for very  high $\mu$-values. One finds   that already  $\mu=128$ gives an excellent approximation of this rotation law. The left panel of Fig. \ref{fig33}   presents essential parts of the instability cones for $\mu=128$ for decreasing magnetic Prandtl numbers in the $(\Ha/\Rey)$ plane. For small $\Pm$ the curves perfectly coincide. The eigensolutions for neutral stability, therefore,  scale with $\Rey$ and $\Ha$ for $\Pm\to 0$.  Hence, the numerical values of
$\Ha$ and $\Rey$ calculated with the inductionless approximation  $\Pm=0$ can also be used for  the
small magnetic Prandtl numbers   of gallium  ($\Pm\simeq 10^{-6}$) or even liquid sodium ($\Pm\simeq 10^{-5}$). This approximation only models the relation $\nu\ll\eta$, it does not mean that $\nu=0$ \citep{R67}. Below we shall demonstrate    that the curves of neutral stability always scale with $\Ha$ and $\Rey$  within this approximation . 
Note also how clear in Fig.  \ref{fig33} (left) the minimum Hartmann number
$\Hamin$ stands at a constant  value for the various small $\Pm$.  

For applications also the growth rates of the instability as the negative imaginary part of the Fourier frequency are relevant. They are   normalized with the rotation rate  $\Om_{\rm out}$ of the outer cylinder  for a fixed Reynolds number and  for almost-stationary inner cylinder by the right panel of Fig. \ref{fig33}. The common Reynolds number  $\Rey=1000$ is  the maximal value in the left panel. By definition the growth rates vanish for the marginal values of the Hartmann number. For $\omega_{\rm gr}/\Om_{\rm out}=1$  the growth time relative  to the rotation time (of the outer cylinder) is $\tau_{\rm gr}/\tau_{\rm rot}=1/2\pi$.  We find  for supercritical magnetic fields  finite values of order 0.2 leading to the typical  relation $\tau_{\rm gr}\simeq \tau_{\rm rot}$.  The $\Pm$-dependence is only weak and  disappears for $\Pm\to 0$. The lines converge for sufficiently small magnetic Prandtl number. Rotation with radially increasing  frequency, therefore,  magnetically decays and grows with  the timescale of rotation itself  - similar to the other forms of magnetorotational instability. 

\begin{figure}
 \centering
  \includegraphics[width=0.75\textwidth]{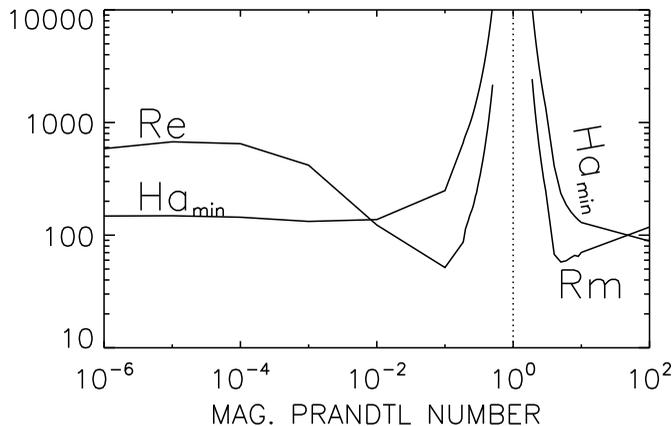}
 \caption{Minimum Hartmann number $\Ha_{\rm min}$  and the corresponding Reynolds numbers  ($\Pm<1$) and magnetic Reynolds numbers ($\Pm>1$) in  dependence on $\Pm$ for   $\mu=128$. 
 For $ \Pm= 1$  
           solutions with positive growth rates do not exist. 
            $\mu_B=\mur=0.9$. Perfect-conducting cylinders.}
 \label{fig3}
\end{figure}

Figure \ref{fig3} gives the  minimum Hartmann number $\Hamin$ together with the associated  Reynolds numbers for strong ($\mu=128$) positive shear in  dependence  on  $\Pm$. One finds   convergency for both eigenvalues   for $\Pm\to 0$.  The
curves basically scale with $\Ha$ and $\Rey$ for $\Pm\to 0$ in both cases. 

For   $\Pm<1$  the Reynolds number takes a minimum at   $\Pm\simeq 0.1$. For even larger $\Pm$ it grows to infinity if $\Pm\to 1$. No solution for $\Pm=1$ exists. For even larger $\Pm$, however, the instability is reanimated  with  Hartmann numbers of the same order as for $\Pm<1$ and rather  low magnetic  Reynolds numbers. Even  for large magnetic Prandtl numbers
the magnetic Mach number remains smaller than unity.  
Instabilities  which only exist for $\nu\neq \eta$ belong to the class of  double-diffusive instabilities \citep{A78}. They do not appear for  $\Pm= 1$ what  possibly means that they do not exist in ideal fluids. \cite{SK15}  derived eigenvalues for rotation laws with positive shear and  for $\Pm=0$ in a short-wave approximation without  boundary conditions, demonstrating the local character of the instability.

In   Fig. \ref{fig3}  the eigenvalues $\Rey$ and $\Ha$ do not depend on $\Pm$ if $\Pm\to 0$. 
The question is whether the global  equation system for the perturbations also possesses  solutions for $\Pm=0$ for all $\mu>1$ or not. We  already know that AMRI solutions with negative shear do exist for $\Pm=0$ but only for flows close to the Rayleigh line with $\mu\leq 0.3$ \citep{HT10}. We ask  whether such a limit also exists for super-rotation. The standard MRI (with axial fields) for all (negative) shear values does not exist for  $\Pm=0$ \citep{C61}. For $\Pm\to 0$  it scales with the magnetic Reynolds number $\Rm$ leading  for small  $\Pm$ to very large Reynolds numbers. This is the explanation  that the experimental realization of the standard MRI is still an existing challenge.

The  dimensionless and linearized equations for the evolution of 
 the  flow, field and pressure  perturbations in the inductionless (or `quasistationary') approximation for $\Pm=0$ are 
\begin{eqnarray}
\Rey \Big(   
\frac{\partial \vec{u}}{\partial t} + ({\vec{ U}}\cdot \nabla)\vec{u}+  
({\vec{u}}\cdot \nabla){\vec{U}}
\Big) =
- \nabla{\it p}+  \Delta \vec{u} 
 +\Ha^2 \Big(  \curl\  \vec{b} \times {\vec{ B}}+\curl\  \vec{ B} \times \vec{b}  \Big)
\label{mom}
\end{eqnarray}
and 
\begin{equation}
\curl(\vec{u\times \vec{B}})+  \Delta\vec{b}=0,
\label{mhd}
\end{equation}
 together with ${\textrm{div}}\ \vec{u} = {\textrm{div}}\ \vec{b} = 0$. 
 The fluctuating and mean magnetic fields are normalized with a  
characteristic scale  $B_0$ of the background field. The background 
flow $\vec U$ is normalized with a characteristic flow 
amplitude $U_0$, the flow perturbations with $\eta/D$ and the  time   with $D/U_0$.   
Obviously, the eigenvalues of the system are $\Rey$ and $\Ha$.  

\begin{figure}
\centering
\includegraphics[width=0.480\textwidth]{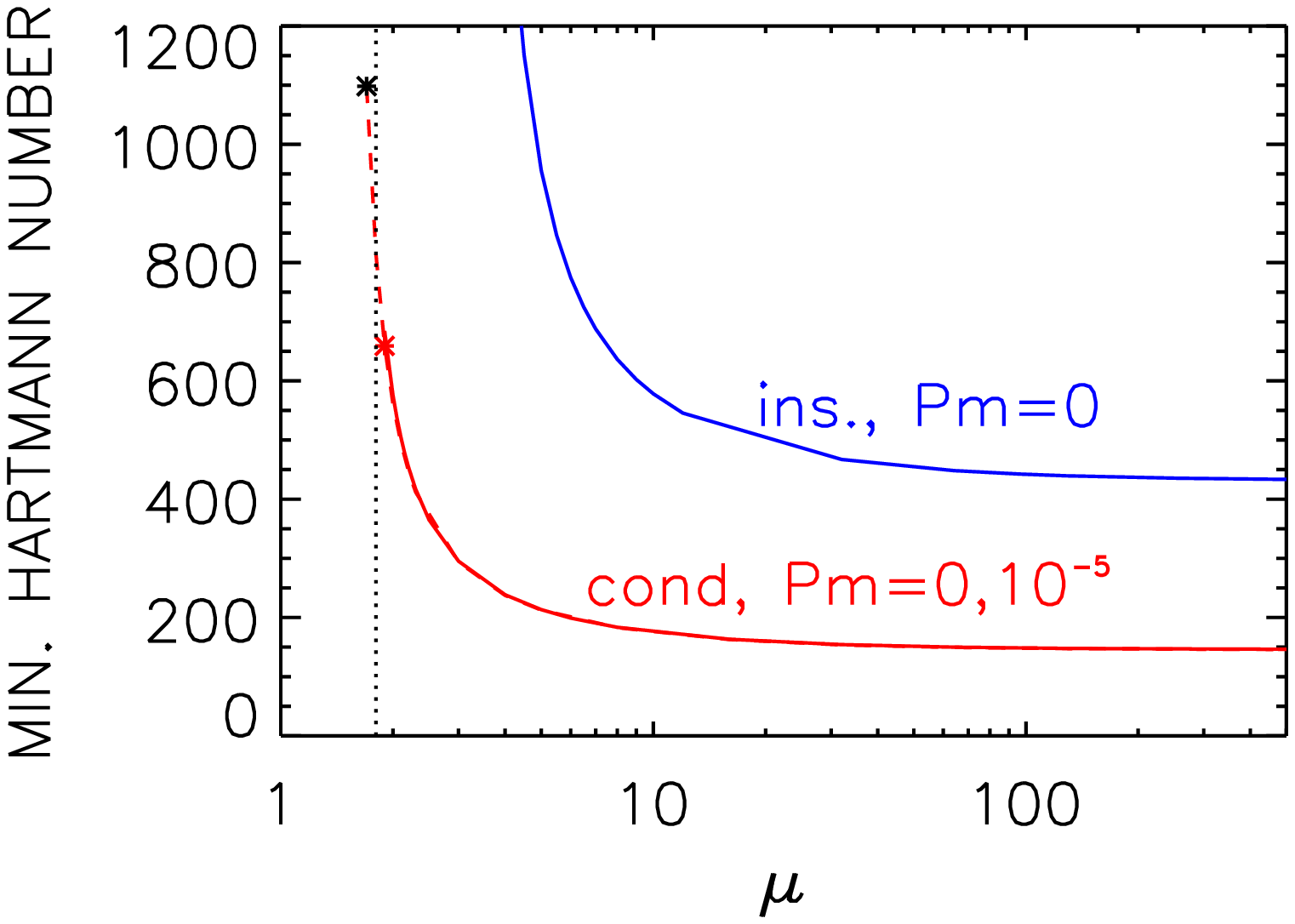}
\includegraphics[width=0.480\textwidth]{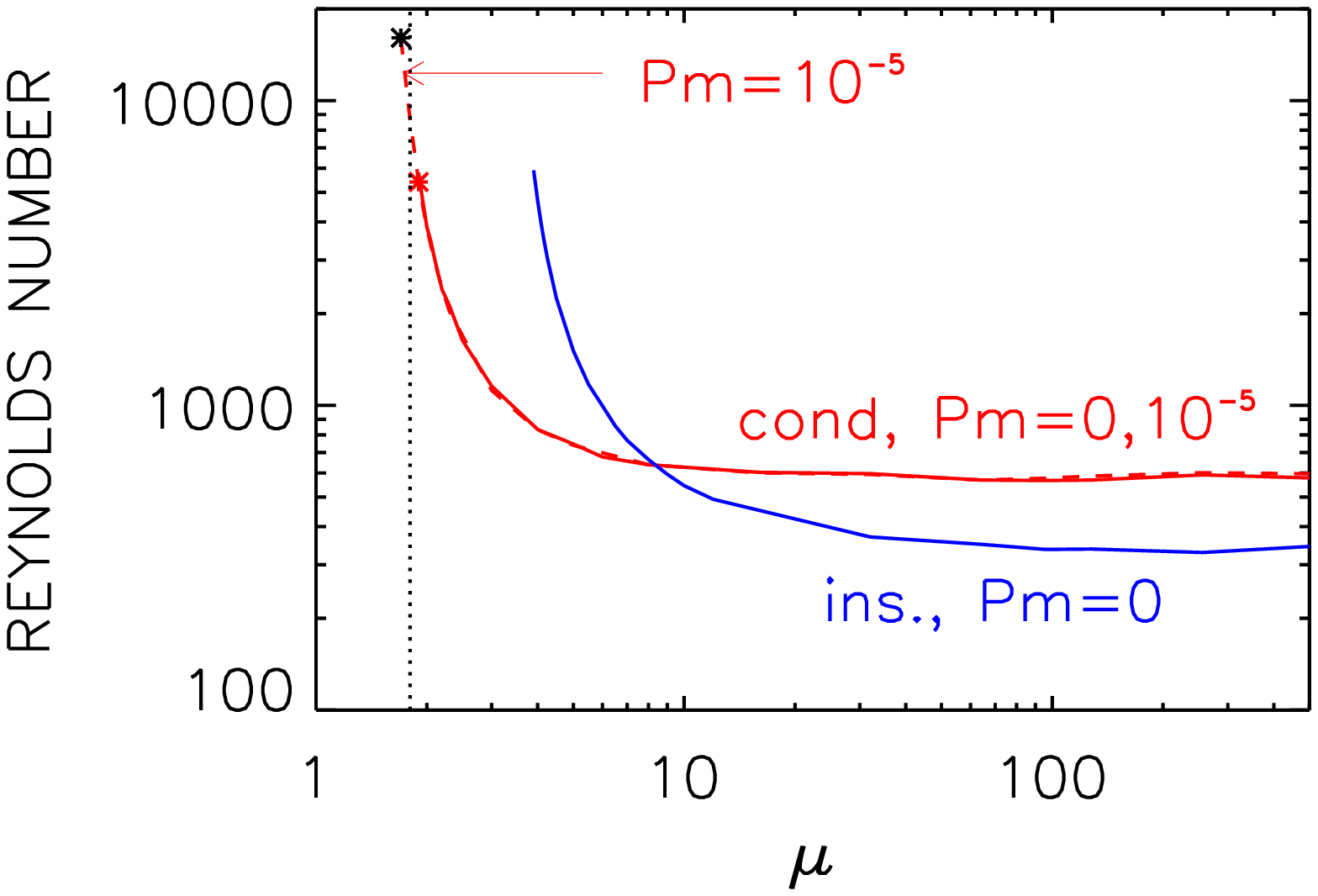}
\caption{$\Hamin$ (left), the corresponding Reynolds numbers (right)  versus the shear number $\mu$  for insulating cylinders (blue, $\Pm=0$) and perfect-conducting cylinders (red, $\Pm=0$, $\Pm=10^{-5}$). The vertical dotted lines mark shear of $\mu_{\rm crit}=1.8$. For $\Pm=0$ no solution for $\mu<1.9$ (red dot) has been found. The dark dot represents the last known solution at $\mu=1.7$ for $\Pm=10^{-5}$. { Note the crossing of  the Reynolds number lines for $\mu\approx 8$}. $m=\pm 1$, $\rin=0.9$.}
 \label{fig3a}
\end{figure}

  The equation system (\ref{mom}) and (\ref{mhd})  is numerically 
 solved with the boundary conditions as described above.
 Figure \ref{fig3a} shows that  for rotation with sufficiently large  positive shear $\mu>\mu_{\rm crit}>1$  solutions for 
$\Pm=0$ indeed  exist for both sorts of  boundary conditions. For perfect-conducting boundaries the dotted vertical line at   $\mu_{\rm crit}=1.8$ appears as a lower bound for the eigensolutions. As indicated  by  black dots, for   $\Pm=10^{-5}$  solutions also exist for $\mu<\mu_{\rm crit}$ but only  for very large  Reynolds and Hartmann numbers. It is  a similar situation as  for negative shear  where solutions for $\mu\gsim 0.3$
only exist for finite $\Pm$ with very  large Reynolds numbers. Note that the lower bounds $\mu_{\rm crit}$ are much harder to extrapolate for insulating boundary conditions than for perfect-conducting ones.

After  Fig.  \ref{fig1}   the global solution with  $\mu=1.9$ and for $\Pm=10^{-5}$  requires much higher Hartmann numbers
and Reynolds numbers than the steeper rotation law with $\mu=2.3$.   
Formally, the critical values for $\mu=1$ (rigid rotation) are  infinite. It is insofar not surprising that both Reynolds numbers and Hartmann numbers are growing for  $\mu\to 1$. For $\Pm=0$, however, no solution for $\mu<1.9$  has been found. 
The curves for $\Pm=0$ indeed suggest the existence of the lower limit  $\mu_{\rm crit}>1$ where the eigenvalues become infinite. Obviously, finite values of the magnetic resistivity   stabilize too flat rotation laws with positive shear in the sense as \cite{LG06} found   for the axisymmetric modes of helical magnetorotational instability within the inductionless short-wave approximation. 

For conducting boundaries also the curves for $\Pm=10^{-5}$ in Fig. \ref{fig3a} are given showing  nearly perfect coincidence with the curve for $\Pm=0$. The two lines hardly can  be distinguished by eyes. The results of calculations in the inductionless approximation can thus be assumed as valid for liquid metals with their small $\Pm$.  Note also that for perfect-conducting cylinders  much weaker magnetic fields than for insulating cylinders are needed for the instability onset. For large values of $\mu$ the ratio of the $\Hamin$  for insulating and perfect-conducting boundaries is about 2.5. The same is true with respect to the Reynolds number but only for small $\mu$ (flat rotation laws)  while for large $\mu$ (steep rotation laws)  the critical Reynolds numbers for vacuum  conditions are { \em smaller} than for conducting boundary conditions. { This behaviour of the eigenvalues  is already known from  standard MRI of quasi-Keplerian rotation laws with axial fields for different boundary conditions}.

\begin{figure}
\centering
\includegraphics[width=0.60\textwidth]{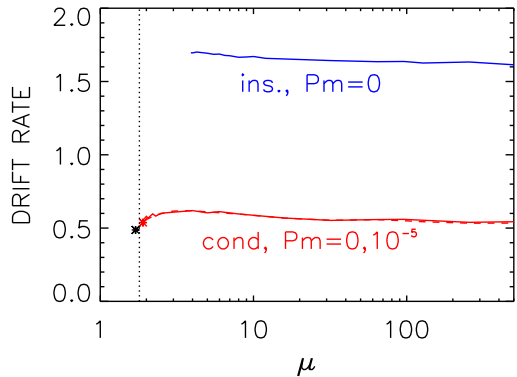}
\caption{Same as in Fig. \ref{fig3a} but for the  drift rate  $\omega_{\rm dr}/\Om_{\rm out}$.}
 \label{fig3b}
\end{figure}

We ask for  the azimuthal drift rates $\omega_{\rm dr}/\Om_{\rm out}$
of the solution with the lowest Hartmann number. In the right panel of Fig. \ref{fig2}  these values for $\Pm=10^{-5}$ only slightly  vary with $\mu$ for $\mu\leq 5$. Figure \ref{fig3b} also shows  the azimuthal drift  measured in terms of  the outer rotation rate  hardly  depending on the shear  or the magnetic Prandtl number but its numerical value basically  depends on the boundary condition. For both boundary conditions and for small $\Pm$ the instability pattern rotates retrograde. The amount of this migration basically depends on the material of the cylinders. The rotation rate of the inner cylinder does not at all  influence the azimuthal migration of the pattern.

\section{Electric currents}\label{sec_rest}
It remains to discuss the influence of the size of the gap between the cylinders on  the excitation conditions and the  axial electric current which produces the minimum  Hartmann numbers. 

\begin{figure}
 \centering
\includegraphics[width=0.70\textwidth]{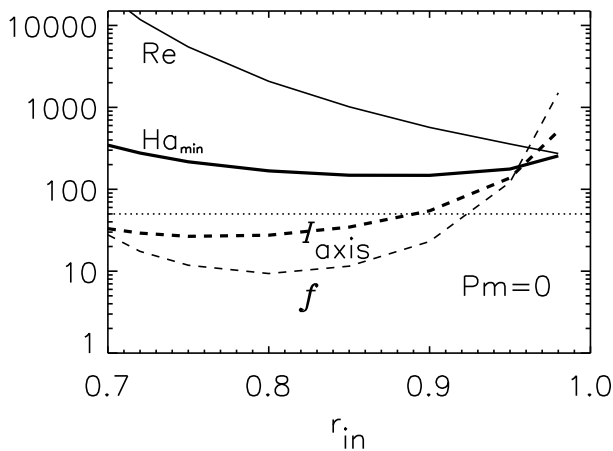}
\caption{$\Hamin$ (thick solid line) and corresponding Reynolds number (thin solid line) versus $\rin$. The necessary  axial electric current $I_{\rm axis}$ (thick dashed line, in kAmp) and rotation frequency $f$ of the outer cylinder (thin dashed line, in Hz) are also given for sodium as the conducting fluid. $R_{\rm out}=5$~cm, $\mu=128$, $m=\pm 1$, $\Pm=0$. Perfect-conducting cylinders.}
 \label{fig41}
\end{figure}

A Taylor-Couette flow with resting inner cylinder is clearly the most simple model for
super-rotating fluids. As it also forms the steepest radial profile one  expects
for this configuration the lowest Hartmann numbers. From Fig. \ref{fig3a} we find, however, that both Reynolds and Hartmann numbers for $\mu\to \infty$ are well represented by the solutions already  for $\mu=128$ and that  perfect-conducting boundaries always lead to the least numerical values.
Figure \ref{fig41} provides the results for variation of the gap width  for  $\Pm=0$  and  for $0.7\leq \rin<1$. The  Hartmann number for   perfect-conducting boundaries  has a minimum  close to $\rin= 0.9$.  In order to transform the values to the axial electric-current 
 within the inner cylinder the relation $I_{\rm axis}=5 R_{\rm in} B_{\rm in}$  
 \citep[see][]{RK13} can 
be written as
\begin{eqnarray}
 I_{\rm axis}= 5 \frac{R_{\rm in}}{D} \ \Ha\ \sqrt{\mu_0\rho\nu\eta},
\label{current} 
\end{eqnarray}
where the axial
currents are measured in Ampere. The numerical values  are given as the thick dashed line in Fig. \ref{fig41}. The  result is that a minimum electric  current   is 26.1 kAmp    for liquid
sodium  as the conducting fluid\footnote{$\sqrt{\mu_0\rho\nu\eta}\simeq 8.2$ in c.g.s.}  with a gap width of 0.25.

The Reynolds numbers associated with the minimal Hartmann numbers given in Fig. \ref{fig41} are moderate enough to ensure laminar flows in nonmagnetic experiments \cite[see][]{B11}.
For stationary  inner cylinders with  $\rin=0.92$ \cite{SG59} found stable solutions for Reynolds numbers not exceeding 40.000 in agreement  with the data of \cite{T36}.
\cite{BC12} detailed discussed the phenomena of subcritical transition to turbulence in 
Taylor-Couette flows by finite disturbances. The experiment described by  \cite{SS14}  to realize the current-free AMRI for negative shear worked with $\rin=0.5$ (with outer radius 8 cm) and a Reynolds number of 2956 without indication of any  nonmagnetic instability within the container. Flows with positive shears are even more stable.

\begin{figure}
 \centering
 \includegraphics[width=0.48\textwidth]{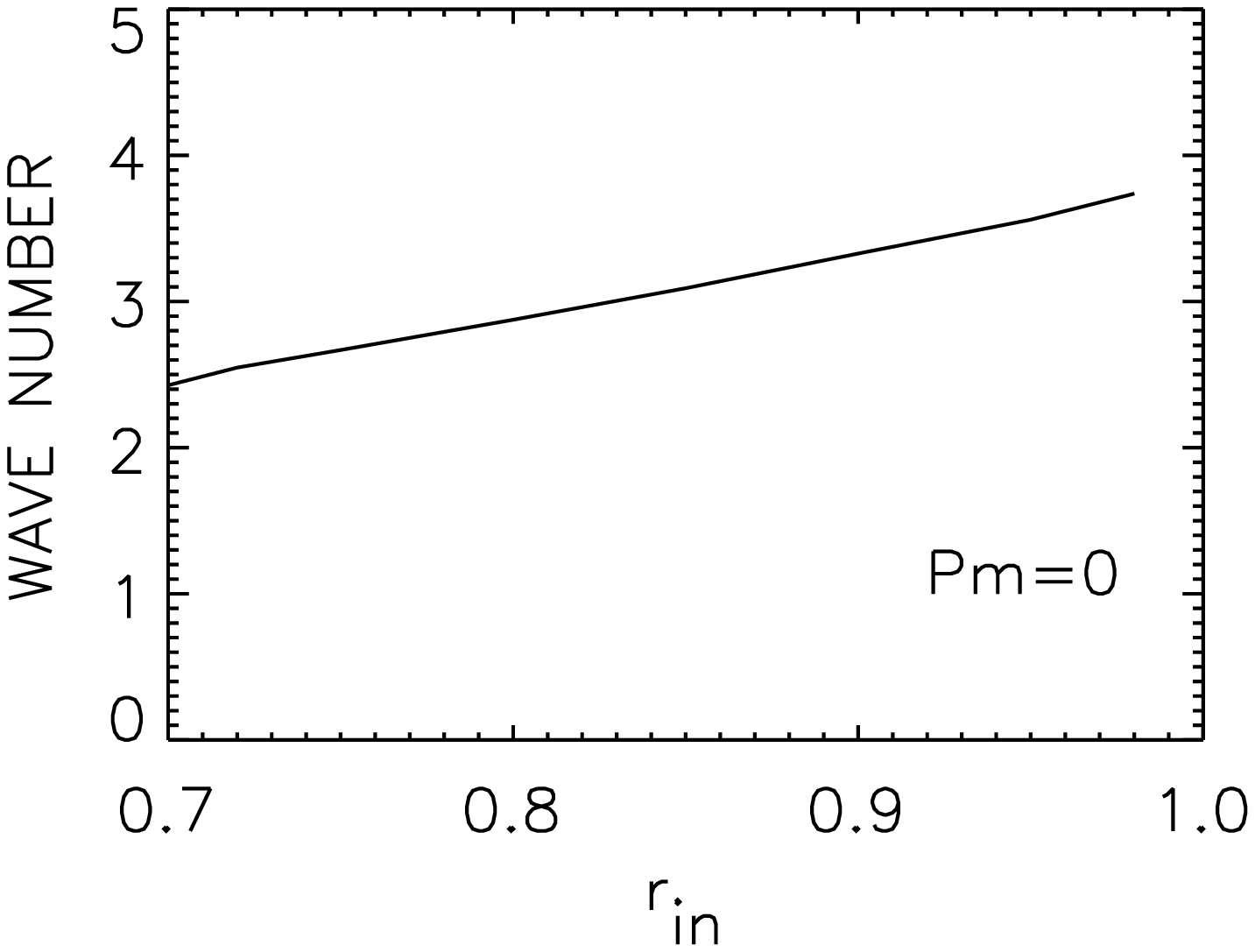}
 \includegraphics[width=0.48\textwidth]{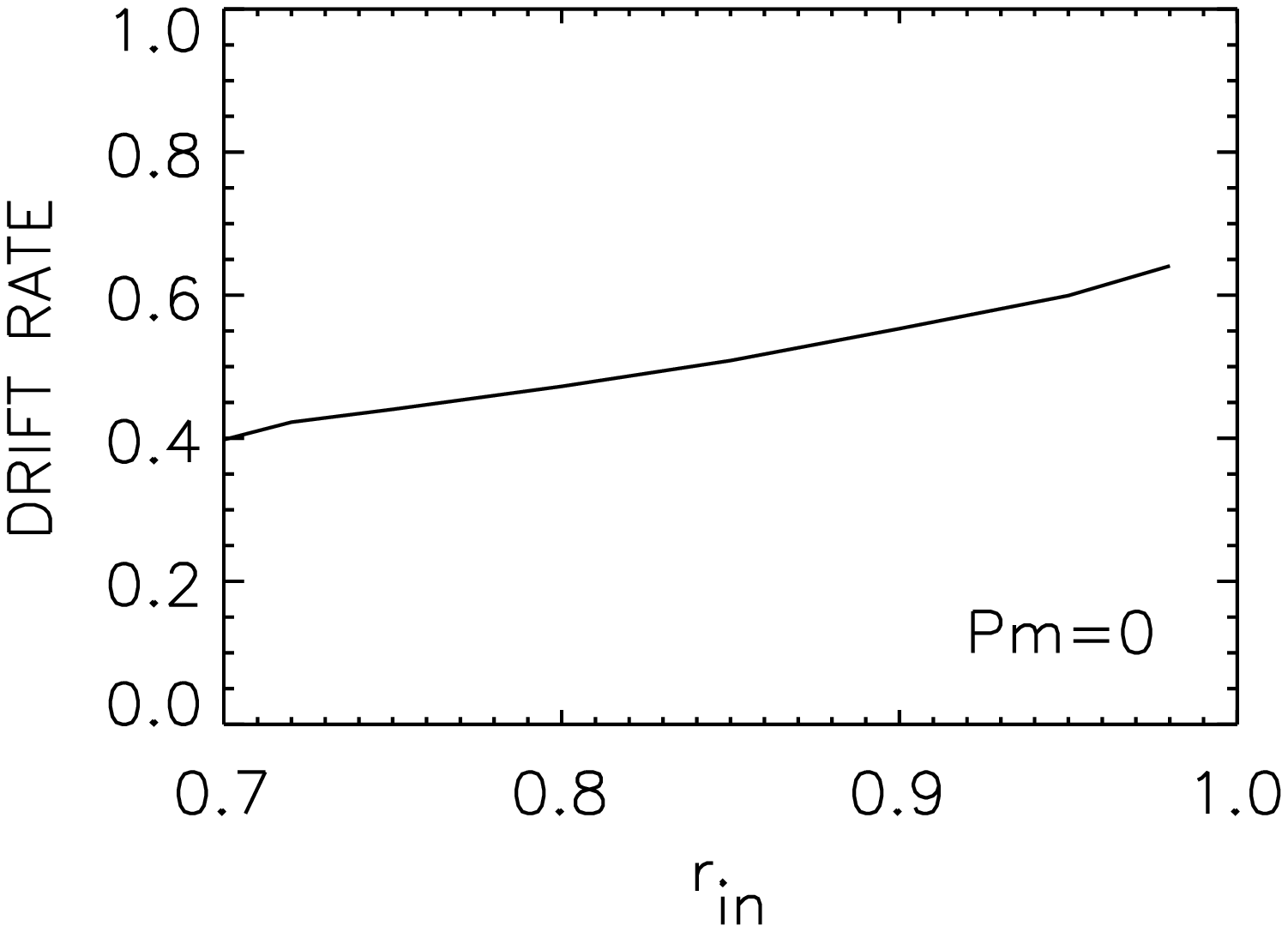} 
 \caption{Same as in Fig. \ref{fig41} but for the  axial wave number $kD$ (left) and the azimuthal drift rate $\omega_{\rm dr}/\Om_{\rm out}$ (right).} 
\label{fig42} 
\end{figure}

 For increasing gap width the Reynolds number continuously grows. However, the rotation frequency $f=\Om_{\rm out}/2\pi$ possesses a minimum at $\rin=0.8$. For liquid sodium    ($\nu=7.1\cdot 10^{-3}$~cm$^2$/s)  and (say) $R_{\rm out}=5$~cm  one obtains  $f=9.3$~Hz.
 For  wider gaps both the critical Reynolds number  
 and Hartmann number   rapidly increase. For $\mur=0.66$ we did not find any instability  assuming $\Rey<15.000$.
Note that  the size of the container does not  influence the needed   electric current  for instability but 
the fractional gap width does. 

The eigenvalues given in Fig. \ref{fig41} are also characterized by  associated axial wave numbers and  azimuthal drift rates. The dependencies of these values on $\rin$  are not strong.  For not too narrow gaps  the cells are hardly  elongated with the rotation axis and the azimuthal migration (\ref{phi})   of the instability pattern is always negative becoming (slightly) slower for wider gaps (Fig. \ref{fig42}).  With the used normalizations both the axial wave number and the drift rates linearly grow with the value of $\rin$; they are maximal for $\rin\to 1$. 
\begin{figure}
 \centering
 \includegraphics[width=0.60\textwidth]{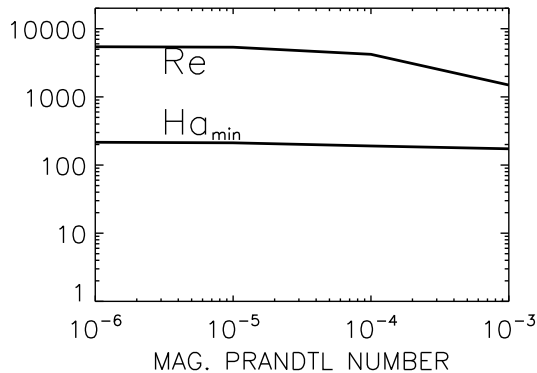}
 \caption{Same as in Fig. \ref{fig3} but for $\rin=0.75$. For $\Pm=10^{-5}$ the Hartmann number is $\Hamin= 211.8$ and the Reynolds number is
	$\Rey = 5360$.  Between  $\Pm\simeq  10^{-3}$ and $\Pm=1$ the values rapidly grow to $\infty$. $\mu=128$,  $\mu_B=\mur=0.75$. Perfect-conducting cylinders.}
 \label{fig43}
\end{figure} 

For the container with $\rin=0.75$,  which after Fig. \ref{fig41} requires the minimum axial electric current for excitation,   the Fig. \ref{fig43} again demonstrates the independence of Hartmann number and Reynolds number of the magnetic Prandtl number. Both values do not depend on $\Pm$ for $\Pm\lsim 10^{-5}$. It is indeed allowed to use for the critical values $\Hamin$ and $\Rey$ the results of the inductionless approximation for liquid metals such as mercury, gallium and sodium as the conducting fluids.  The magnetic Mach number (\ref{mach})  for the largest $\Pm$ in Fig. \ref{fig43} is of order unity but for  smaller $\Pm$ -- as for all instabilities which scale with $\Ha$ and $\Rey$ for $\Pm\to 0$ -- the magnetic Mach number is much smaller than unity. 


After Eq.  (\ref{current}) the electric currents needed for neutral stability  behave like  the numerical values of the material expression $\sqrt{\mu_0\rho\nu\eta}$ which mainly runs with $\sqrt{\rho}$ as for the mentioned liquid metals the averaged diffusivity  $\sqrt{\nu\eta}\approx$~const.  It is thus clear that  experiments for liquid sodium  require the weakest electric currents.

Sofar we did not vary the radial profile of the magnetic background field. One of the 
questions is how sensitive the described instability  reacts on
the presence of axial electric-currents {\em within} the fluid. The Tayler instability  of the  $z$-pinch with $B_\phi\propto R$ -- which for $\Ha=\Ha_0$ is unstable against nonaxisymmetric perturbations for all $\Pm$  even without rotation -- is supported by super-rotation unless $\Pm=1$ so that even subcritical Hartmann numbers $\Ha<\Ha_0$ exist \citep{RS16}.  
 Similar to our above results the azimuthal migration  of the nonaxisymmetric  instability pattern  depends on $\Pm$. Also here  the pattern  counter-rotates  for $\Pm \ll 1$ and it co-rotates for $\Pm\gg 1$.
\begin{figure}
\centering
\includegraphics[width=0.60\textwidth]{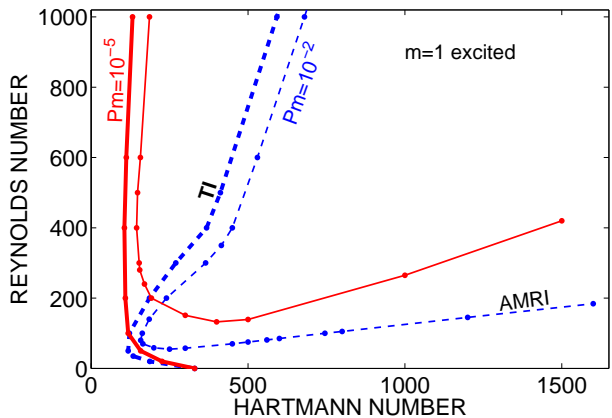} 
\caption{Stability map for resting inner cylinder. The radial profiles of the  background fields vary from $\mu_B=1/\rin$ 
(TI, thick lines) to $\mu_B=\rin$ (AMRI, thin lines) and for  $\Pm=0.02$ (blue) and  $\Pm=10^{-5}$ (red). $\Ha_0=332$ for all $\Pm$. Note the coincidence of the upper branches  for large Reynolds numbers.  $\rin=0.9$,  perfect-conducting boundaries.} 
\label{fig2a} 
\end{figure} 
 
In addition for the current-free field $B_\phi\propto 1/R$  Fig. \ref{fig2a} also yields the stability  lines for the field $B_\phi\propto R$  for two different magnetic Prandtl numbers. The instability for $\Rey=0$ appears at  $\Ha_0$  \citep{T57,T73,V72,SS12} which   does not depend on  $\Pm$  \citep{RS10}.
For large Reynolds numbers  only small differences of the lines of marginal instability exist for the two  radial profiles.   The minimum Hartmann number for instability are slightly  reduced by the electric currents, $\Hamin<\Ha_0$, but the differences are small. The main  result is that the magnetic-induced instability of super-rotating  Taylor-Couette  flows with narrow gaps is only slightly influenced by  the applied radial field profile.  The upper branches  in Fig.  \ref{fig2a}  demonstrate this phenomenon for two different magnetic Prandtl numbers. One can  even consider the field as uniform between the cylinders \citep{E58}.  The magnetic field only acts as a catalyst for the instability. 

\begin{figure}
\centering
\hbox{
\includegraphics[width=0.5\textwidth]{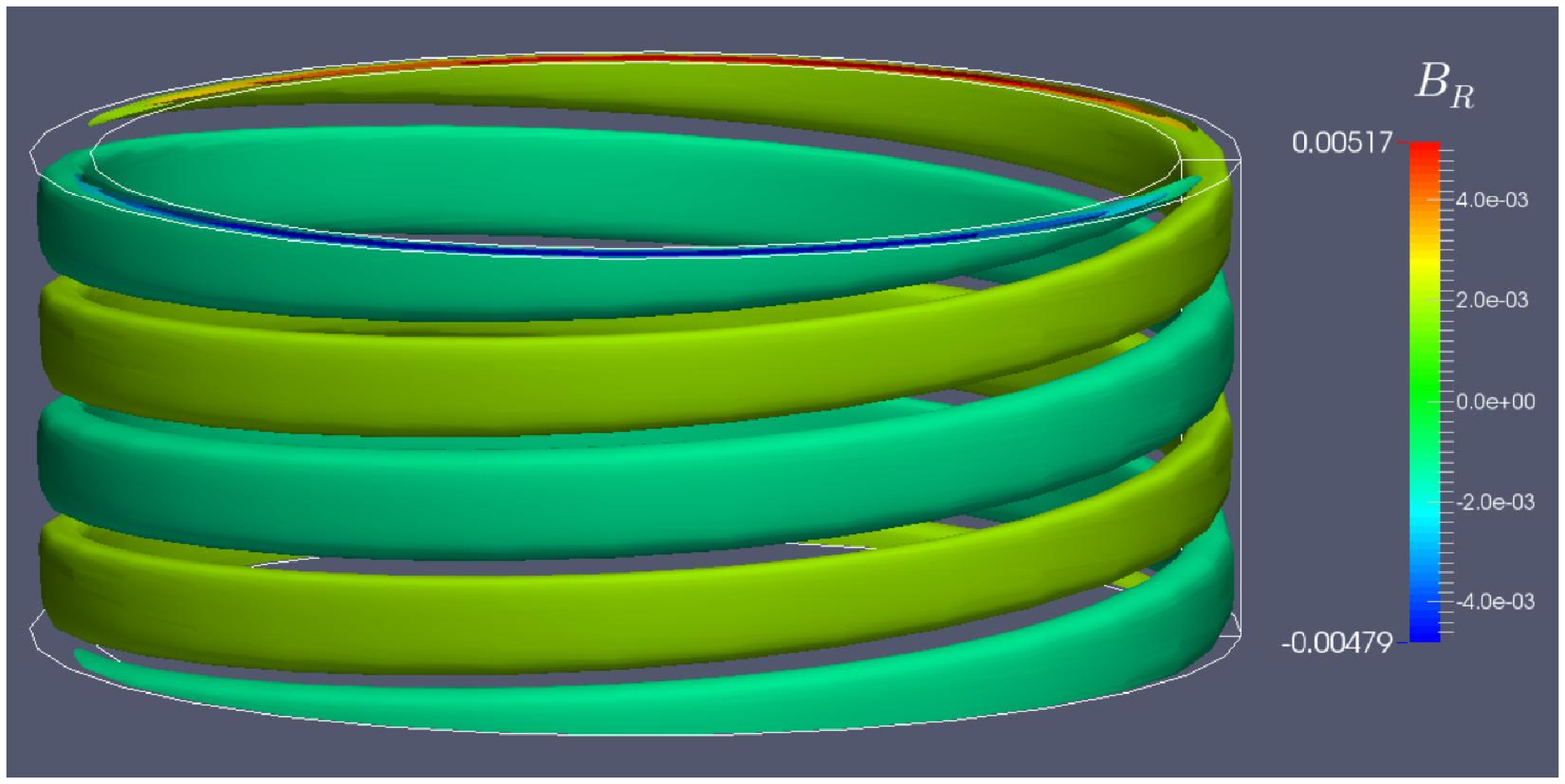} 
\includegraphics[width=0.5\textwidth]{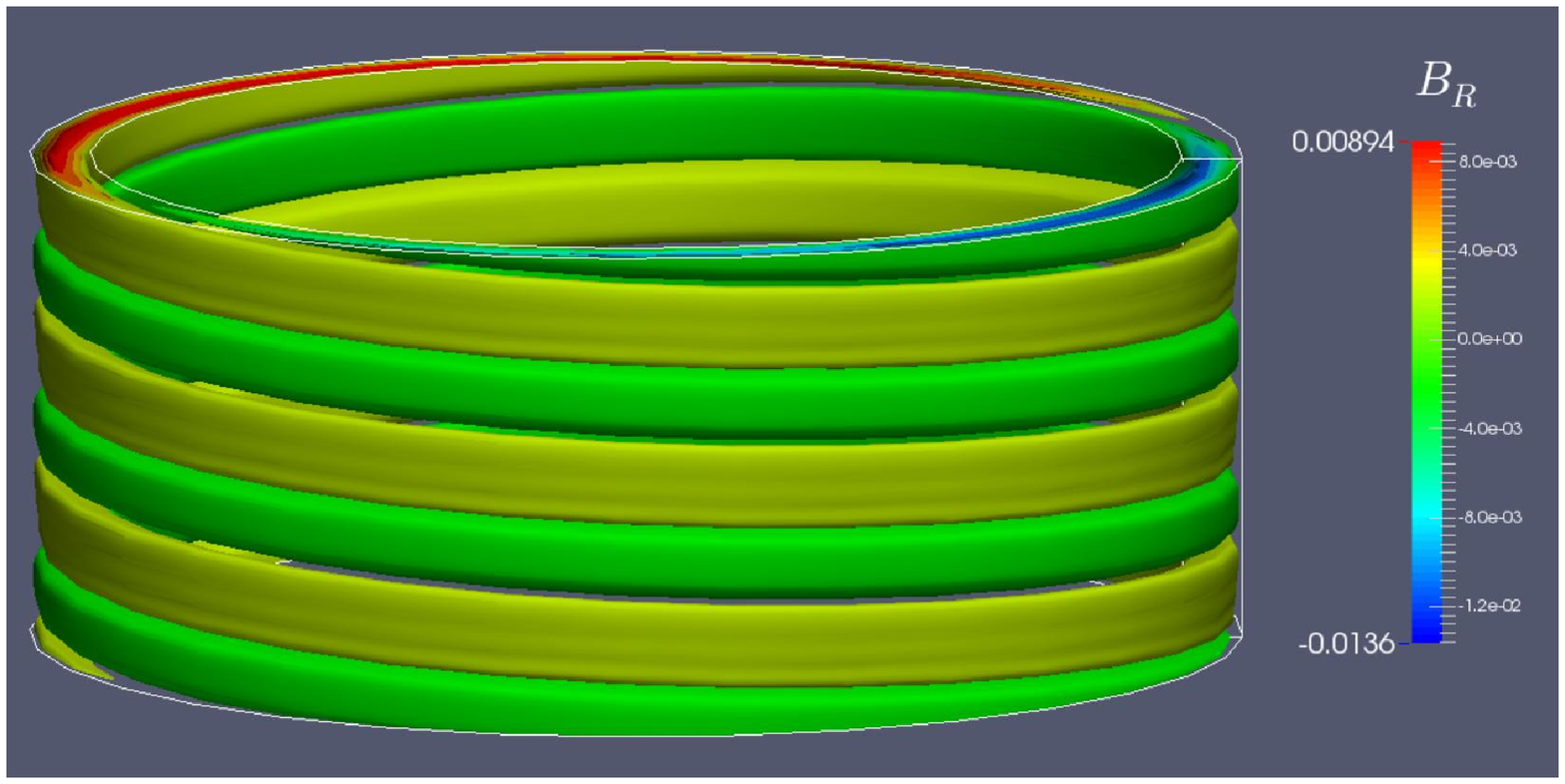} }
\caption{{ Isolines of the radial magnetic field component for instability without axial electric current between the cylinders ($\mu_B=\rin$, left) and with axial electric current  ($\mu_B=1/\rin$, right). Resting inner cylinder.  $\Ha=300$, $\rin=0.9$, $\Rey=155$, $\Pm=0.02$, perfect-conducting boundaries.}} 
\label{fig4a} 
\end{figure}

{ To visualize the shape of the instability pattern a {\em nonlinear} spectral  code is used which has been developed from the hydrodynamic code of  described by \cite{FB04} and works with the expansion of the solution after azimuthal Fourier modes.
Figure \ref{fig4a} demonstrates that  the resulting patterns of the instabilities for resting inner cylinder and for $\Pm=0.02$ only sightly differ in dependence of the existence of axial electric  background currents.} The  isolines of the radial magnetic field component are given showing  different helicities of the two cases for the given sign of the electric current) and   the axial wavelength in the left panel exceeds that of the  instability model in the right panel. 

 The two examples  in Fig. \ref{fig4a} have identical supercritical Reynolds number and Hartmann number. In the right panel a uniform axial electric current exists in the fluid between the cylinders so that the pattern belongs to the Tayler instability (under the influence of differential rotation). The electric current is missing in the left panel but the cell structure is very similar. For  strong super-rotation the two instabilities appear to  fuse and both instabilities are hard to distinguish unless $\Pm=1$. 

\section{Conclusions}
In an earlier paper we have shown that  Taylor-Couette flows with radially increasing rotation  can become unstable against nonaxisymmetric perturbations  under the presence of sufficiently strong azimuthal magnetic fields which are due to a homogeneous electric current \citep{RS16}. 
The present paper leads to the conclusion that such instability also exists  if the field between the cylinders is current-free. At least for small $\Pm$ under the influence of differential rotation  the differences between both instabilities   become very small. Both  instabilities   appear for $\Pm\neq 1$ while   the sign of the azimuthal pattern drift strongly differs for $\Pm\ll1$ and $  \Pm\gg1$. For $\Pm=1$ rotation with positive shear   stabilizes  TI and  completely suppresses the instability of current-free fields. Solutions for AMRI  only exist if the molecular viscosity and the microscopic magnetic diffusivity strongly differ, hence  solution do not exist for $\Pm= 1$ which is the  characteristics of double-diffusive instabilities. 

 


An important difference to the standard magnetorotational instability with axial fields (which  
 only exists for negative shear) is that the new  instability also appears in the inductionless approximation with $\Pm=0$ for sufficiently large positive shear. The numerical calculations suggest a lower limit  of $\mu_{\rm crit}=1.8$. For finite but small magnetic Prandtl numbers solutions exist even below this limit but with very large Reynolds numbers.
 The consequence is that for finite but small magnetic Prandtl numbers the solutions for $\mu>\mu_{\rm crit}$ scale with $\Rey$ and $\Ha$ almost independent of  $\Pm$. This gives the chance to probe the numerical results by experiments in the laboratory by use of liquid metals with their extremely low magnetic Prandtl numbers. For the optimal  gap width of 0.25 and for liquid sodium as the conducting fluid the minimal axial current is 26.1 kAmp and the rotation frequency is about 10 Hz  (for $R_{\rm out}=5$~cm). These numbers should be suitable for experiments. However, in view of the significant differences of $\Hamin$ for perfect-conducting and insulating boundaries (Fig. \ref{fig3a}) any real experiment with sodium  flowing between (say) copper walls remains to be carefully predicted.

\bibliographystyle{jpp}
\bibliography{superamri.bib}

\end{document}